\documentclass[12pt, draftclsnofoot, onecolumn]{IEEEtran}

\input epsf
\usepackage{graphicx,epstopdf}   
\usepackage{pifont,dsfont}
\usepackage{bm,bbm}
\usepackage{subcaption}
\usepackage{float}
\usepackage[utf8x]{inputenc}
\usepackage{color}
\usepackage{algorithmic,mathtools}
\usepackage[table]{xcolor}
\definecolor{Gray}{gray}{0.9}

\usepackage{amssymb,amsmath,amsfonts,amsthm}
\usepackage{cite,citesort}
\usepackage{balance}
\usepackage[utf8x]{inputenc}
\usepackage{graphicx}
\usepackage{epsfig}

\usepackage[
top    = 1.70cm,
bottom = 1.05in,
left   = 0.60 in,
right  = 0.60 in]{geometry}

\IEEEoverridecommandlockouts

\newtheorem{theorem}{Theorem}

\newtheorem{corollary}{Corollary}

\begin{document}

\title{Online Supervised Learning for Traffic Load Prediction in Framed-ALOHA Networks}

\author{
\IEEEauthorblockN{Nan Jiang, Yansha Deng, Osvaldo Simeone, and Arumugam Nallanathan}\\

\thanks{ This work was supported by the Engineering and Physical Sciences Research Council (EPSRC), U.K., under Grant EP/R006466/1 and the European Research Council (ERC) under the European Union Horizon 2020 research and innovative programme (grant agreement No 725731).}
\thanks{
N. Jiang, and A. Nallanathan are with School of Electronic Engineering and Computer Science, Queen Mary University of London, London, UK (e-mail: \{nan.jiang, a.nallanathan\}@qmul.ac.uk).
}
\thanks{
Y. Deng, and O. Simeone are with Department of Informatics, King's College London, London, UK (e-mail: \{yansha.deng, osvaldo.simeone\} @kcl.ac.uk). 
}

\vspace*{-0.6cm}
}



\maketitle

\begin{abstract}

Predicting the current backlog, or traffic load, in framed-ALOHA networks enables the optimization of resource allocation, e.g., of the frame size. However, this prediction is made difficult by the lack of information about the cardinality of collisions and by possibly complex packet generation statistics. Assuming no prior information about the traffic model, apart from a bound on its temporal memory, this paper develops an online learning-based adaptive traffic load prediction method that is based on Recurrent Neural Networks (RNN) and specifically on the Long Short-Term Memory (LSTM) architecture. In order to enable online training in the absence of feedback on the exact cardinality of collisions, the proposed strategy leverages a novel approximate labeling technique that is inspired by Method of Moments (MOM) estimators. Numerical results show that the proposed online predictor considerably outperforms conventional methods and is able to adapt to changing traffic statistics.

\end{abstract}
\vspace*{-0.1cm}

\vspace*{-0.2cm}
\section{Introduction}
\vspace*{-0.0cm}
Framed-ALOHA (f-ALOHA) has been widely adopted as a key component of multiple access protocols in many state-of-the-art wireless communication systems, including Long-Term Evolution (LTE) and 5G New Radio (NR). In f-ALOHA, time is organized into time frames, with each frame containing multiple Random Access Opportunities (RAOs). RAOs refer to subsets of channel resources in time, frequency, or/and code domain, e.g., random access preambles in the LTE system. In each frame, devices select RAOs at random and transmit to the connected Base Station (BS) in an uncoordinated manner. Collisions cause devices to retransmit in following frames, increasing the backlog of packets to be transmitted in a frame beyond the load due to newly generated packets. Adapting the number of RAOs per frame to the estimated current backlog, or traffic load, is an important step to relieve network congestion and reduce access delays. However, prediction is made difficult by the fact that the BS does not have access to the cardinality of the collisions, that is, to the number of devices that have selected the same RAO, and by the possibly complex nature of the incoming traffic statistics (see Fig. 1). As an example, the incoming traffic may consist of mixtures of different traffic types, including periodic, event-driven (bursty), multimedia streaming patterns, and etc. \cite{shafiq2012first,3GPP2011Study}.

\captionsetup{singlelinecheck=false} 
\begin{figure}[t]
\vspace*{-0.0cm}
    \begin{center}
        \begin{minipage}[t]{0.6\textwidth}
    \centering
        \includegraphics[width=1\textwidth]{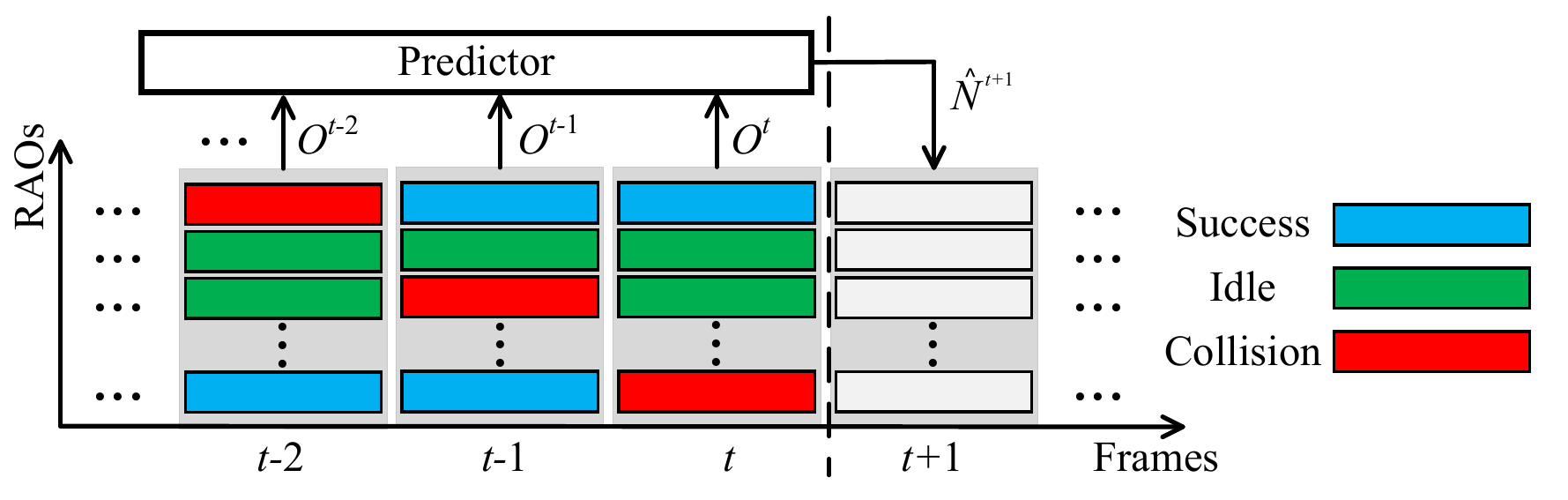}
        \vspace*{-0.5cm}
        \caption{\scriptsize Timeline of an f-ALOHA protocol and target predictor based on historical data about collided, successful, and idle RAOs in previous frames.}
                \label{fig1}
        \end{minipage}
    \end{center}
    \vspace*{-0.8cm}
\end{figure}

Estimating traffic backlog in f-ALOHA can only rely on the observation on the number of RAOs in each frame that are idle, collided, or successful (see Fig. 1). Previous classical works have proposed Method of Moment (MOM)-based estimators that aim at matching the average number of such RAOs to the current measurements \cite{schoute1983dynamic}. More recent works proposed to predict bursty traffic for event-driven applications, i.e., for massive devices being activated by an external event to request transmissions within a short period, using drift analysis \cite{wu2012fast}, MOM \cite{duan2016d}, or Maximum-Likelihood (ML) estimation \cite{he2017traffic}. All these prior works estimate the current backlog only based on the latest observations of idle, collided, or successful RAOs, while ignoring historical data from prior frames. As argued in this work, information about RAOs in previous frames can be useful to learn features of the traffic statistics that enable an improved prediction.

In this work, we target modern Internet of Things (IoT) traffic scenarios with complex statistics, possibly encompassing mixtures of long- and short-memory processes, e.g., a mixture of random and periodic transmissions with long duty cycles \cite{shafiq2012first,3GPP2011Study}. In order to capture the complex dynamics of the IoT traffic, we propose an online supervised learning method that adopts a Recurrent Neural Network (RNN) model based on the state-of-the-art Long Short-Term Memory (LSTM) architecture \cite{hochreiter1997long}. The most relevant prior work is \cite{aceto2019mobile}, where the authors leverage LSTM to classify mobile encrypted traffic. To the best of our knowledge, the application of LSTM to traffic backlog prediction in f-ALOHA has not been considered before. The proposed method is able to adapt at runtime by leveraging a novel supervised technique based on approximate labeling. In particular, given the absence of feedback on the exact cardinality of collisions, approximate target labels defining the current backlog are estimated in a manner inspired by the discussed memoryless MOM solutions \cite{schoute1983dynamic,duan2016d}.

The rest of the paper is organized as follows. Section II describes the system model and problem formulation. Section III provides background material. Section IV introduces the proposed online supervised learning method. Section V presents numerical results, and Section VI concludes the paper.

\vspace*{-0.1cm}
\section{System Model and Problem Formulation}
\vspace*{-0.05cm}
We consider a f-ALOHA network system consisting of an arbitrary number of devices and a single BS. As illustrated in Fig. \ref{fig1}, time is divided into frames, where each frame contains $F$ RAOs. At the given frame $t$, a number of devices $N^t$ are \emph{active}, that is, they either have tried unsuccessfully to access previous frames or they have newly generated packets to transmit. The number $N^t$ represents the current \emph{backlog} of the system, also referred to as \emph{traffic load}. We assume an arbitrary process for the generation of new packets, which is unknown to the BS. Every device has only two possible states, either inactive or active, while it has a single packet to be transmitted in the latter case. This captures sporadic traffic random access in cellular-based networks as detailed in \cite{3GPP2011Study,he2017traffic,duan2016d,wu2012fast}.

Each active device chooses one of the $F$ RAOs uniformly at random for transmission. As a result of the random selections of the RAOs, a transmission may fail if a collision occurs, i.e., two or more devices select the same RAO at the same frame. For each RAO, the BS can detect whether it contains a single transmission, a collision, or whether it is idle, i.e., it contains no transmitted packet. To take into account the effects of radio channels, e.g., path-loss, fading, and interference, we further assume that an RAO that contains a collision or a single transmission can be erroneously detected as idle with probability $p_\text{ed}$ \cite{3GPP2011Study}. If a packet is successful detected without collision, the corresponding transmitting device receives an acknowledgement message from the BS. If a packet is either collided or undetected, it is re-transmitted on a randomly selected RAO in the next frame unless a maximum number $\gamma_{\text max}$ of attempts is exceeded. These re-tranmissions result in backlog accumulation in future frames, which can in turn increase the collision probability. The processing latency of sequence transmission and acknowledgement are ignored as in prior works \cite{3GPP2011Study,schoute1983dynamic,he2017traffic,duan2016d,wu2012fast}.

In this letter, the objective is to predict the number $N^{t+1}$ of devices that are active at the beginning of each frame $t+1$. To do so, the BS keeps a record of the observations $O^{t'}$ made in prior and current frames $t' = 1$, $2$, $\cdots$, $t$. Each observation $O^{t'}=$ $\{{V}^{t'}_\text{i},{V}^{t'}_\text{s},{V}^{t'}_\text{c}\}$ includes the number ${V}^{t'}_\text{i}$ of idle RAOs, the number ${V}^{t'}_\text{s}$ of successful detected RAOs, and the number ${V}^{t'}_\text{c}$ of collided RAOs, as detected by the BS (possibly in an erroneous manner, as discussed). Using the observed history $H^t=\{O^1,O^2,\cdots,O^{t}\}$ in frame $t$, the goal of this paper is to predict the forthcoming traffic value $N^{t+1}$ by learning a conditional distribution ${\mathbb P} \left\{ N^{t+1} =  n |H^t \right\}$ and then solving the Maximum A Posteriori (MAP) problem
\vspace*{-0.1cm}
\begin{align}\label{q1}
&(\text{P1}): {\hat{N}}^{t+1} = \mathop {\text{arg max}}\limits_{ n\in\{0,1,...,N_\text{max}\}}     \quad  {\mathbb P} \left\{ N^{t+1} =  n |H^t \right\},
\end{align}
where the $N_\text{max}$ is an upper bound on the backlog. Unless one makes simplistic assumptions as in \cite{wu2012fast,he2017traffic,duan2016d}, problem (1) remains generally intractable even in the presence of realistic traffic models, further justifying the data-driven solutions developed in prior art and in this letter. We emphasize that we focus solely on the problem of prediction and that we leave the problem of investigating the interplay between overload control via, e.g., frame size selection and access barring, and traffic prediction to future work.

\vspace*{-0.1cm}
\section{Conventional Backlog Prediction}
\vspace*{-0.0cm}
In this section, we review two conventional traffic prediction methods, namely MOM-based algorithms \cite{schoute1983dynamic} (see also \cite[Sec. III]{jiang2019cooperative,jiang2019deep}\cite[Sec. V-A]{duan2016d}), and the current state-of-the-art ML estimator of \cite{he2017traffic}. Both these methods use an estimate $\tilde{N}^t$ for the backlog $N^{t}$ in the current frame, given observations in frame $t$, as the prediction $\hat{N}^{t+1}$ for the backlog in frame $t+1$. We emphasize that, throughout this letter, given information $H^t$ available at the end of frame $t$, we use the notation $\tilde{N}^t$ to represent an estimate of the backlog in frame $t$, while $\hat{N}^{t+1}$ denotes a prediction for the backlog at the beginning of frame $t+1$.

\vspace*{-0.1cm}
\subsection{MOM-Based Estimator}\label{sec3a}
\vspace*{-0.0cm}
Given a backlog $N^t=n$ in any frame $t$, neglecting the possibility of detection errors, the expected number of RAOs in idle, success, and collisions state are given respectively as
\vspace*{-0.1cm}
\begin{align}\label{q6-0}
&E_\text{i}(n)  =  {\mathbb E} [ {V}_\text{i}  | N^t = n] = F\bigg(1-\frac{1}{F}\bigg)^n,
 \\\label{q6-1}
&E_\text{s}(n)  =  {\mathbb E} [ {V}_\text{s}  | N^t = n] = n\bigg(1-\frac{1}{F}\bigg)^{n-1},
 \\\label{q6-2}
&E_\text{c}(n)  =  {\mathbb E} [ {V}_\text{c}  | N^t = n] = F\left(\vphantom{\left(1-\frac{1}{F}\right)^{n-1}}  1- \left(1-\frac{1}{F}\right)^{n}    - \frac{n}{F} \left(1-\frac{1}{F}\right)^{n-1} \right),
\end{align}
where we recall that $F$ is the number of RAOs. These expectations can be easily computed by noting that each active of the $n$ devices selects any of the $F$ RAOs with equal probability. 

MOM estimators $\tilde{N}_\text{MOM}^t$ of the current backlog $N^{t}$ aim at matching one or more of the moments in (\ref{q6-0}), (\ref{q6-1}), and (\ref{q6-2}) to the current observations ${V}^{t}_\text{i}$, ${V}^{t}_\text{s}$, and ${V}^{t}_\text{c}$, respectively. A MOM estimator hence generally finds a value of $n$ that minimizes a measure of the discrepancy between the moments in (\ref{q6-0}), (\ref{q6-1}), and (\ref{q6-2}), on the one hand, and the respective observations, on the other. For instance, one could consider the Mean Absolute Error (MAE) 
\begin{align}\label{q7}
\varphi^{t}(n) =  \frac{1}{3}(|E_\text{i}(n) - V^{t}_\text{i}|  + |E_\text{s}(n) - V^{t}_\text{s}| + |E_\text{c}(n) - V^{t}_\text{c}|),
\end{align} which would yield the MOM-based estimator\begin{equation}\label{eq:MOM} \hat{N}_\text{MOM}^{t+1} = \tilde{N}_\text{MOM}^{t} = \mathop {\text{arg min}}\limits_{ n \in \{0,1,...,N_\text{max}\}}   \varphi^{t}(n). \end{equation}

Simplified, and generally less accurate, MOM-based estimators that enjoy closed-form solutions have been proposed in \cite{duan2016d,jiang2019cooperative}. As an example, in \cite[Sec. 3]{jiang2019cooperative}, traffic load $N^{t}$ is estimated by matching the single moment (\ref{q6-0}) with the given observation ${V}^{t}_\text{i}$. Imposing the equality $E_\text{i}(n)={V}^{t}_\text{i}$ and using the rounded solution $n$ as the estimate $\tilde{N}^{t+1}$ of the backlog for the current frame yields the estimator
\vspace*{-0.1cm}
\begin{align}\label{q3}
\hat{N}_\text{MOM}^{t+1}  = \tilde{N}_\text{MOM}^{t} = \text{round} \left\{ \text{log}_{ (1-\frac{1}{F})}\left(\frac{{V}^{t}_\text{i}}{F}\right) \right\},
\end{align}
where $\text{round}\{\cdot\}$ is the nearest integer function.

\subsection{ML Estimator}\label{sec3b}
A more complex estimator can be obtained by using the ML estimator $\tilde{N}_\text{ML}^{t}$ of the current backlog \cite{he2017traffic}. This is given as 
\vspace*{-0.1cm}
\begin{align}\label{q3-1}
 \hat{N}_\text{ML}^{t+1}= \tilde{N}_\text{ML}^{t} = \mathop {\text{arg max}}\limits_{ n\in\{0,1,...,N_{\text{max}}\}}     \quad  {\mathbb P} \{ O^{t} | N^{t} =  n  \}.
\end{align}
Solving problem (\ref{q3-1}) requires the computation of the probability ${\mathbb P} \{ O^{t} | N^{t} =  n  \}$ for each possible $n\in\{0,1,...,N_{\text{max}}\}$ given the current observation $O^t$. Note that each value ${\mathbb P} \{ O^{t} | N^{t} =  n  \}$ represents the likelihood of a value $n$ given the current observation. Reference \cite{he2017traffic} proposes a numerical approach that computes the vector of probabilities, or likelihoods, ${\mathbb P} \{ O^{t} | N^{t} =  n  \}$ for all $n\in\{0,1,...,N_{\text{max}}\}$ as the steady-state probability vector of a Markov chain. This Markov chain is defined by letting each device sequentially, and independently, select an RAO at each step. We refer to \cite{he2017traffic} for details on the numerical procedure.

\vspace*{-0.0cm}
\section{Online Supervised Learning-Based Backlog Prediction}\label{SIV}
\vspace*{-0.0cm}

In this section, we propose an online supervised learning approach for the training of a predictor of the forthcoming traffic load ${N}^{t+1}$ given the observations $H^t$ available at the end of frame $t$. Unlike the existing methods presented above, the proposed scheme aims at capturing not only the information present in the most recent observation $O^t$, but also the historical information in the previous observations in $H^t$ in order to detect patterns in the traffic generation mechanism and in the communication protocol. To this end, the method leverages an RNN, and specifically an LSTM architecture, which is capable of recognizing patterns in temporal data.

\captionsetup{singlelinecheck=false} 
\begin{figure}[t]
\vspace*{-0.0cm}
    \begin{center}
        \begin{minipage}[t]{0.6\textwidth}
    \centering
        \includegraphics[width=1\textwidth]{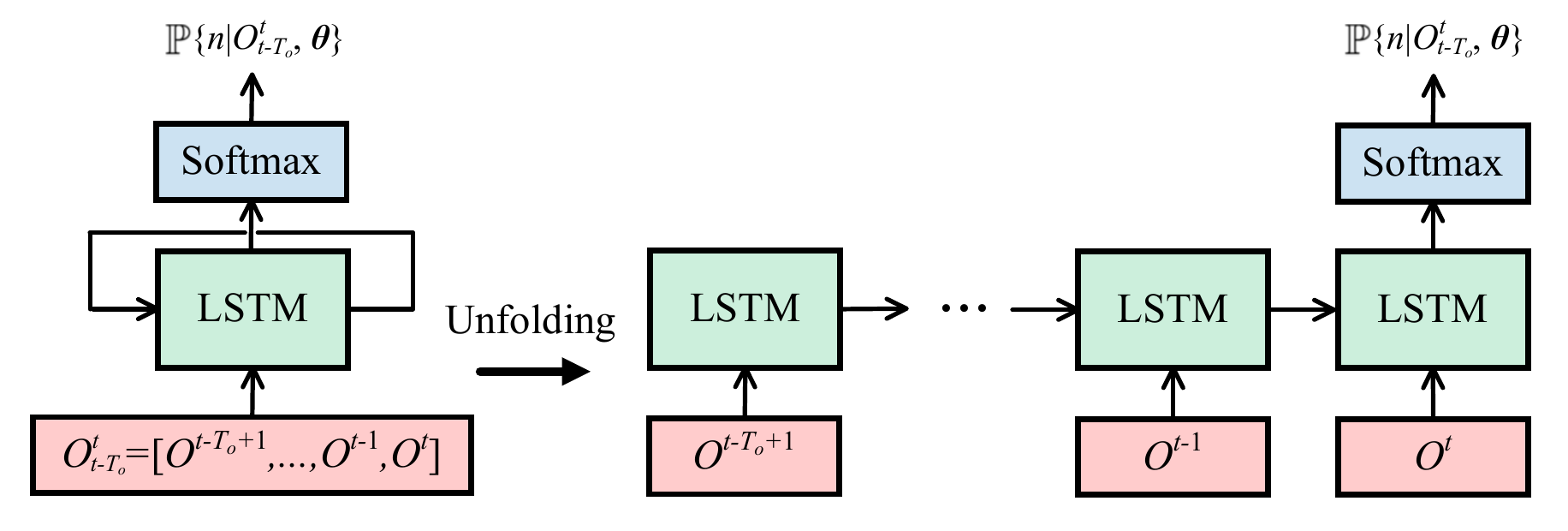}
        \vspace*{-0.5cm}
        \caption{\scriptsize Proposed LSTM architecture (left) with temporal unfolding (right). The LSTM block can contain multiple stacked LSTM layers that are fully connected.}
                \label{fig2}
        \end{minipage}
    \end{center}
    \vspace*{-0.6cm}
\end{figure}

As illustrated in Fig. \ref{fig2}, the RNN is sequentially fed the observation $O^t$ and is tasked with the goal of reproducing $\hat{N}^{t+1}$ at any time $t$. Since the latter value is never observed at the BS, the proposed method substitutes it with an estimator $\tilde{N}^{t+1}$ obtained in the next frame $t+1$ using one of the methods described in Sec. III. In this way, with a delay of one frame, the weights of the RNN are adapted in order to minimize the error of the predictor $\hat{N}^{t+1}$ output by the RNN at frame $t$ with respect to the estimate $\tilde{N}^{t+1}$ computed in the following frame. Note that this estimate uses information not available in frame $t$ and hence it allows the RNN to observe and detect patterns in the history $H^t$ that correlate with future traffic loads. Details are provided next.

The proposed LSTM architecture is illustrated in Fig. \ref{fig2}. An LSTM layer, consisting of multiple standard LSTM units \cite{hochreiter1997long} receives at each frame $t$ the current observation vector $O^t$ and is connected to an output layer consisting of a softmax non-linearity with $N_\text{max}$+1 output values. The softmax layer outputs an estimate ${\mathbb P} \{ \hat{N}^{t+1} = n | O_{t-T_o}^t, \bm{\theta}  \}$ of the probability of each possible value $n$ for the forthcoming backlog $N^{t+1}$ given a window $T_o$ of previous observations $O_{t-T_o}^t=[O^{t-T_o+1},\cdots,O^{t-1},O^{t}]$. The probability ${\mathbb P} \{ \hat{N}^{t+1} = n | O_{t-T_o}^t, \bm{\theta}  \}$ is parameterized by the vector $\theta$ that contains both the LSTM internal parameters and the weights of the softmax layer. 

More precisely, we adopt a \emph{stateless} implementation of LSTM whereby, as shown in the right-hand side of Fig. \ref{fig2}, at each frame $t$ the LSTM network's network state is re-initialized and the network is progressively fed the observations $O^{t-T_o+1},\cdots,O^{t-1},$ and $O^{t}$, producing the vector of probabilities ${\mathbb P} \{ \hat{N}^{t+1} = n | O_{t-T_o}^t, \bm{\theta}  \}$ upon observing vector $O_{t-T_o}^t$. Parameter $T_o$ should generally be selected so as to capture the expected memory for the packet generation process and communication protocol. Examples will be given in the next section. We note that stateless implementations are generally known to be easier to train and more stable than their stateful counterparts, in which the state of the LSTM layer is not re-initialized at each step \cite{zaremba2014recurrent}.

In order to adapt the model parameter $\theta$, we adopt standard stochastic gradient descent implemented via BackPropagation Through Time (BPTT) \cite{werbos1990backpropagation}. In particular, at each frame $t+1$, the BS estimates the current backlog $\tilde{N}^{t+1}$ using one of the methods discussed in Sec. III. Then, it updates the weights $\theta$ in the direction of the negative gradient of the cross-entropy loss \vspace*{-0.1cm}
\begin{align}\label{q5}
 L^{t}({\bm{\theta}}) =   - \text{log}\left({\mathbb P} \{ \hat{N}^{t+1} = \tilde{N}^{t+1} | O_{t-T_o}^{t+1}, \bm{\theta}  \}\right),
\end{align} where we recall that the probability ${\mathbb P} \{ \hat{N}^{t+1} = \tilde{N}^{t+1} | O_{t-T_o}^{t+1}, \bm{\theta}  \}$ is defined by the LSTM and by the softmax layer. The gradient can be computed via BPTT using standard methods. In practice, rather than applying the gradient of Eq. (\ref{q5}) at frame $t+1$, it is preferable to consider a window, or random mini-batch, of $T_b$ previous values and evaluate the gradient of the average loss 
\vspace*{-0.1cm}
\begin{align}\label{q5-1}
 L^{t}({\bm{\theta}}) =   - \sum_{t'=t-T_b+1}^{t} \text{log}\left({\mathbb P} \{ \hat{N}^{t'+1} = \tilde{N}^{t'+1} | O_{t'-T_o}^{t'+1}, \bm{\theta}  \}\right).
\end{align} This can generally reduce the variance of the stochastic gradient and improve stability of training \cite{bishop2006pattern}.

In order to reduce the time and computational resource needed for convergence of LSTM training, it is useful to initialize the weights of the LSTM by first running offline experiments based on available traffic models, which can be mismatched to online traffic statistics. This may be considered as an example of meta-learning \cite{park2019learning}. We will provide an example in the next section.

\vspace*{-0.2cm}
\section{Numerical Results}
\vspace*{-0.0cm}

In this section, numerical experiments are conducted to evaluate the traffic load prediction accuracy of the proposed online supervised learning method. We mostly assume the presence of $N_\text{u}=1000$ devices generating a packet in any frame independently with probability $0.005$, as well as of additional $N_\text{p}=20$ devices generating one packet every $T_\text{p}=10$ frames in a deterministic (periodic) manner. This scenario captures the coexistence of services with both random and periodic traffic types \cite{shafiq2012first,3GPP2011Study}. We will also consider random bursty traffic \cite[Ch. 6.1]{3GPP2011Study}, as detailed latter. Unless stated otherwise, other parameters are set according to 3GPP technical report for Machine-Type Communication \cite{3GPP2011Study} as follows: error detection probability $p_\text{ed}=0.05$; retransmission constraint $\gamma_\text{max}=10$; and number of RAOs $F=54$.

\captionsetup{singlelinecheck=true}
\begin{table}[htbp!]
\setcounter{table}{0}
	\centering
	\caption{Supervised Learning Hyperparameters \vspace*{-0.0cm}}
	{\renewcommand{\arraystretch}{0.6}
		\begin{tabular}{|*{1}{p{5cm}}|*{1}{p{2cm}} |}
			\hline
			\rowcolor{Gray}
		   \bf{Hyperparameters}   &    \bf{Value}$\vphantom{\Big(}$  \\  \hline
		   $\vphantom{\big(}$Memory $T_o$ & 20 \\
		   $\vphantom{\big(}$RMSProp learning rate $\alpha$ & 0.0001 \\
		   $\vphantom{\big(}$LSTM drop-out rate & 0.2 \\
           $\vphantom{\big(}$Minibatch size $\vphantom{\big(}$ & 64 \\
           $\vphantom{\big(}$Historical samples size $T_\text{b}$ & 1000 \\
                        \hline
		\end{tabular}
	}
	\vspace*{-0.1cm}
	\label{table1}
\end{table}

We compare the performance in terms of prediction error among the MOM predictor (\ref{q3}), the ML predictor (\ref{q3-1}), and three LSTM-based predictors. The first, referred to as ``Offline LSTM", trains the LSTM during an offline phase with $10^5$ frames of synthetically generated traffic. For the offline phase, the exact backlog $N^{t+1}$ can be used when training using the cross-entropy criterion. The statistics of the offline traffic are different from the online traffic in that the former only contains the underlying random traffic, while the latter also models periodic traffic. This offline scheme is treated as the baseline, and its weights are transferred to the online LSTM-based predictors as initialization. The second scheme, referred to as ``Online LSTM", implements the proposed online scheme by using the MOM estimator (\ref{q7}) in the cross-entropy criterion (\ref{q5-1}), while adapting to the online traffic statistics. The third, referred to as ``Genie-aided LSTM", trains the LSTM by using criterion Eq. (10) with the correct value $N^{t+1}$. The performance of this scheme provides an upper bound, due to its use of an ideal supervision in the form of signal $N^{t+1}$.

\captionsetup{singlelinecheck=false} 
\begin{figure}[htbp!]
\vspace*{-0.4cm}
    \begin{center}
        \begin{minipage}[t]{0.6\textwidth}
    \centering
        \includegraphics[width=1\textwidth]
 {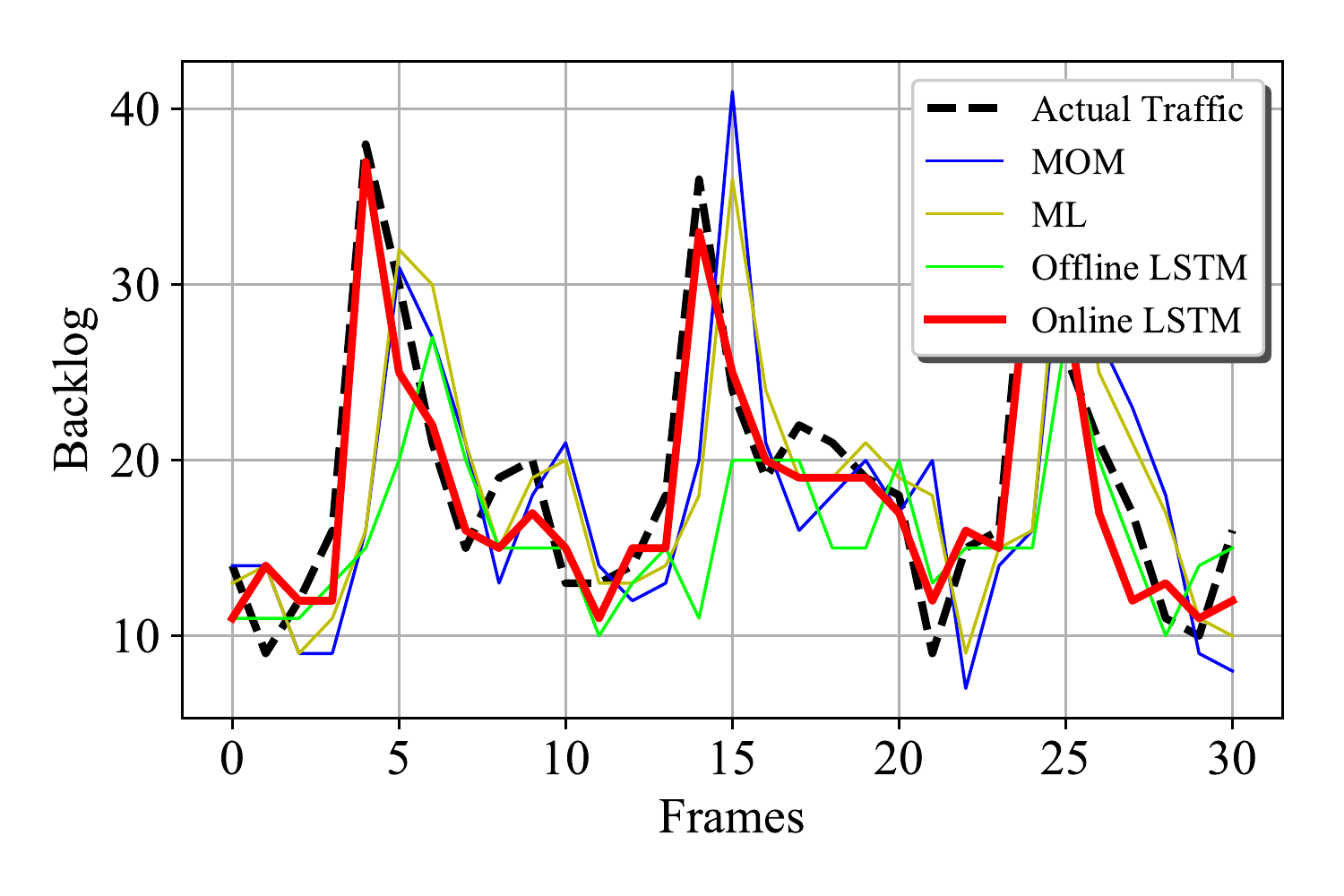}
        \vspace*{-0.8cm}
        \caption{\scriptsize Actual traffic and predicted backlog versus the number of frames after 10$^5$ online training frames.}               \label{fig3}
        \end{minipage}
    \end{center}
    \vspace*{-0.6cm}
\end{figure}

We start by illustrating the operation of the proposed and reference predictors in Fig. \ref{fig3}. This figure plots the actual and predicted backlog after 10$^5$ frames along time frames. We observe that the only method that is able to predict the backlog spikes due to periodic traffic is the proposed Online LSTM method (the Genie-aided LSTM scheme is not shown). In fact, both MOM and ML are not capable of capturing historical trends in the traffic, and Offline LSTM has only observed data without packets from the periodic traffic.

\begin{figure}[htbp!]
\vspace*{-0.4cm}
    \begin{center}
        \begin{minipage}[t]{0.6\textwidth}
    \centering
        \includegraphics[width=1\textwidth]{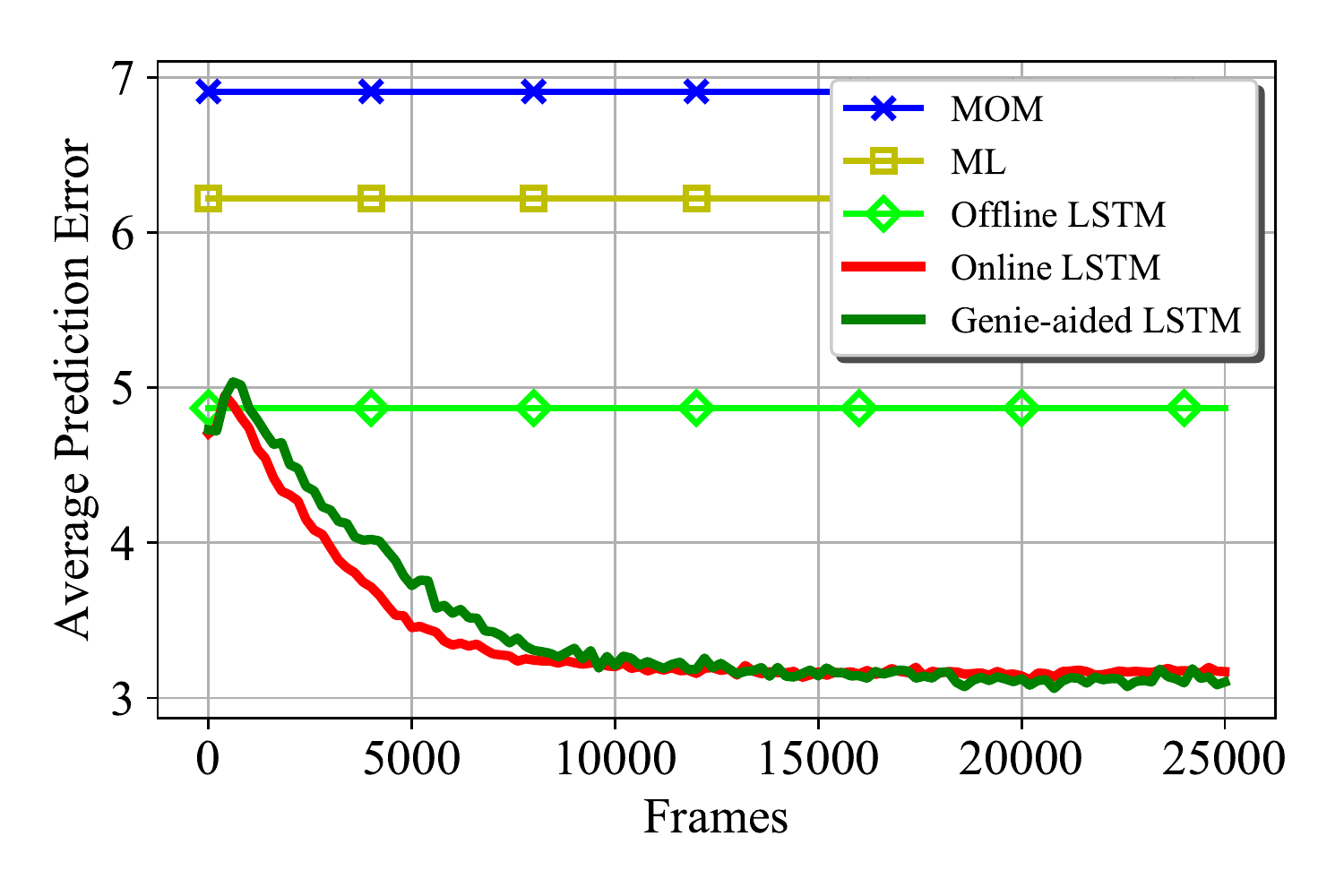}
        \vspace*{-0.9cm}
        \caption{\scriptsize Average prediction error per frame as a function of the frames in the online adaptation phase.}
                \label{fig4}
        \end{minipage}
    \end{center}
    \vspace*{-0.5cm}
\end{figure}

Fig. \ref{fig4} shows the evolution (averaged over 500 training trails) of the absolute prediction error $|\hat{N}^{t+1}-{N}^{t+1}|$ as a function of the frames observed in the online phase for the proposed online LSTM scheme and the genie-aided LSTM scheme. It is seen that the proposed online scheme is able to fairly quickly adapt to the traffic conditions, improving over the performance of MOM and ML strategies and converging to those of the ideal genie-aided scheme. It is also noted that the LSTM scheme outperforms MOM and ML even with the initialized weights obtained from the offline training phase based only on random traffic.

\captionsetup{singlelinecheck=false} 

\captionsetup{singlelinecheck=false} 
\begin{figure}[htbp!]
    \begin{center}
    \begin{minipage}[t]{0.6\textwidth}
    \centering
        \includegraphics[width=1\textwidth]{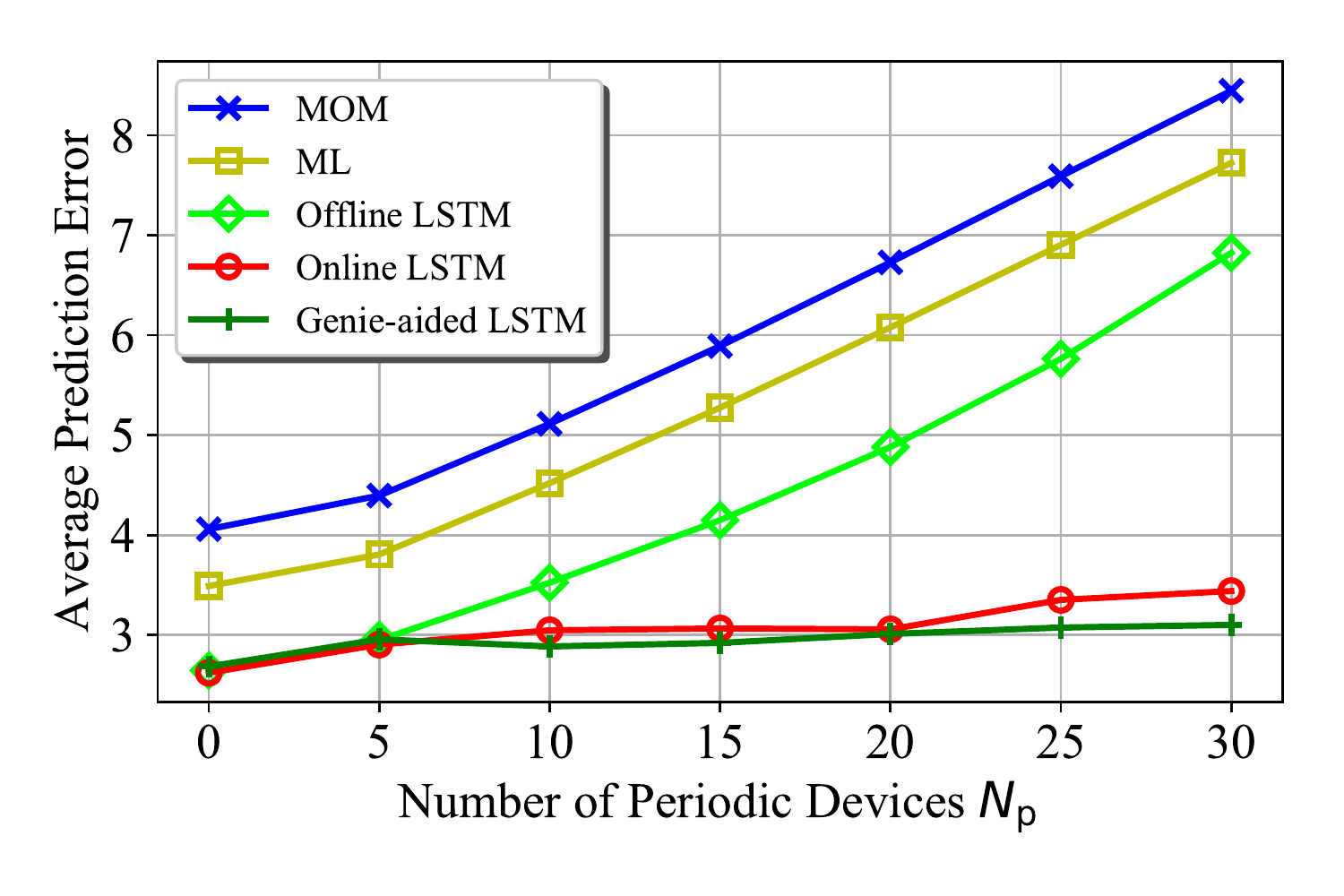}
        \vspace*{-0.8cm}
        \caption{\scriptsize Average prediction errors per frame versus the number of devices with periodic traffic.}
            \label{fig5}
    \end{minipage}
    \end{center}
\vspace*{-0.5cm}
\end{figure}

Fig. \ref{fig5} plots the prediction error versus different number $N_\text{p}$ of devices with periodic traffic. It can be seen that the ideal Genie-aided LSTM scheme slightly outperforms the proposed Online LSTM scheme, while their performance gap is small. Apart from Genie-aided LSTM, we observe that increasing $N_\text{p}$ degrades the prediction accuracy of all methods, but the proposed online method demonstrates a significantly smaller degradation as compared to the other strategies. This is due to the capability of the proposed scheme to adapt to the traffic statistics. It is also interesting to note that, even in the absence of statistical regularities in the traffic, i.e., with $N_\text{p}=0$, the LSTM-based schemes outperform MOM and ML. This is also because the random access communication protocol itself presents temporal correlations due to retransmissions of the same devices after collisions.

\begin{figure}[htbp!]
    \begin{center}
        \begin{minipage}[t]{0.6\textwidth}
    \centering
        \includegraphics[width=1\textwidth]{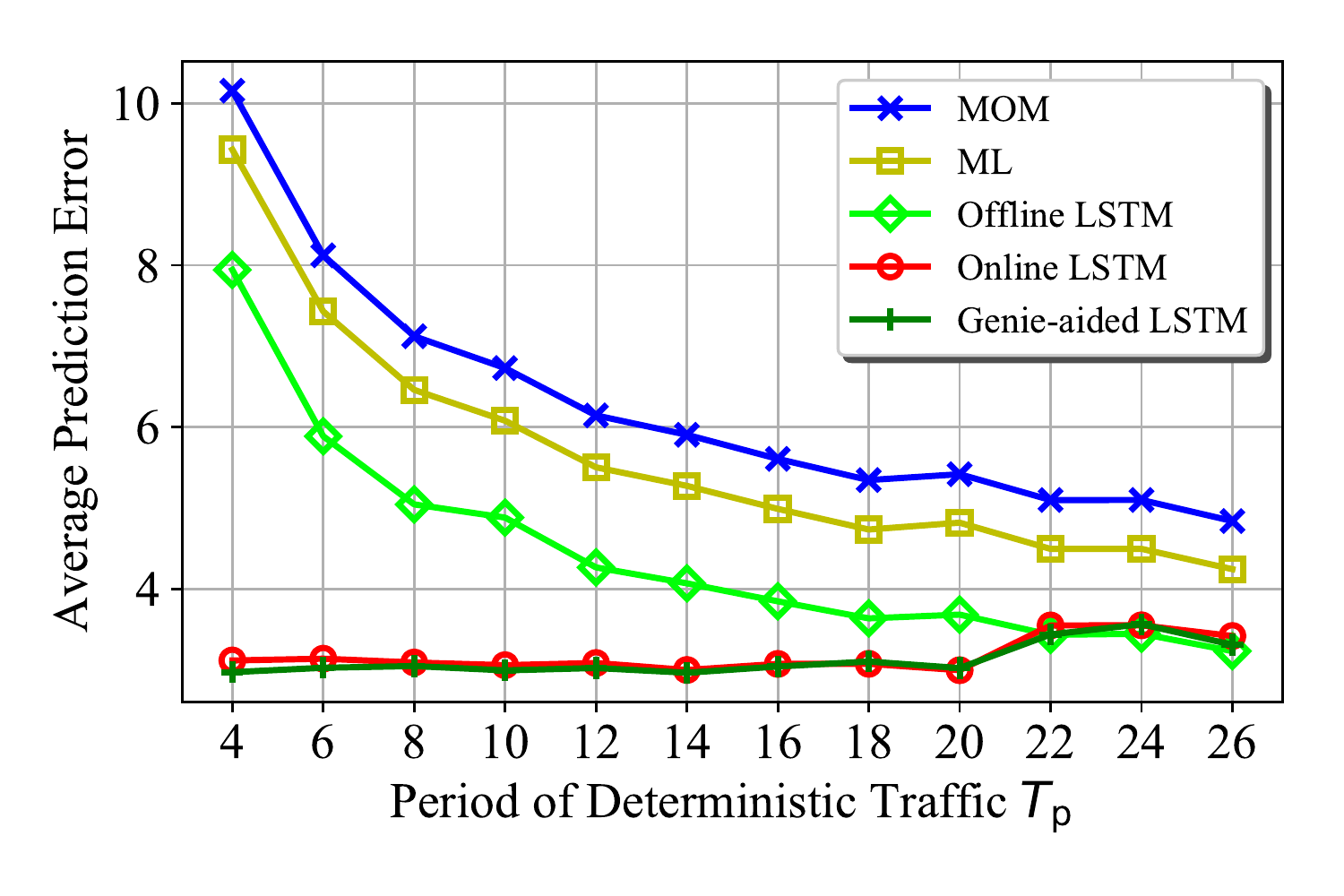}
        \vspace*{-0.8cm}
        \caption{\scriptsize Average prediction errors per frame versus the packet generation period of deterministic traffic.}
                \label{fig6}
        \end{minipage}
    \end{center}
\vspace*{-0.5cm}
\end{figure}

Fig. \ref{fig6} shows the prediction error as a function of the packet generation period $T_{\text{p}}$ of the periodic traffic. Increasing the period $T_\text{p}$ results in a smaller traffic load, which improves the prediction accuracy of MOM, ML, and Offline LSTM. In contrast, as long as $T_\text{p}$ is not too large, the prediction accuracy of the proposed LSTM is not significantly affected by the change in $T_\text{p}$ due to its capability to adapt to the traffic statistics. Specifically, this is only true if $T_\text{p}\le T_o=20$, that is, if the traffic periodicity is smaller than the memory of the LSTM predictor. In contrast, when $T_\text{p}>20$, the prediction accuracy of LSTM suddenly degrades to the same level of Offline LSTM scheme. This is because, with a memory equal to $T_o=20$, the LSTM predictor cannot capture any traffic correlation pertaining to frames occurring more than 20 frames before the frame of interest. Therefore, if $T_\text{p}>20$, online adaptation cannot improve the prediction accuracy. This degradation can be eliminated by increasing the memory of LSTM $T_o$, but at the cost of increasing the required computational and data resources for both training and prediction.

\captionsetup{singlelinecheck=false} 

\begin{figure}[htbp!]
    \begin{center}
    \begin{minipage}[t]{0.6\textwidth}
    \centering
        \includegraphics[width=1\textwidth]{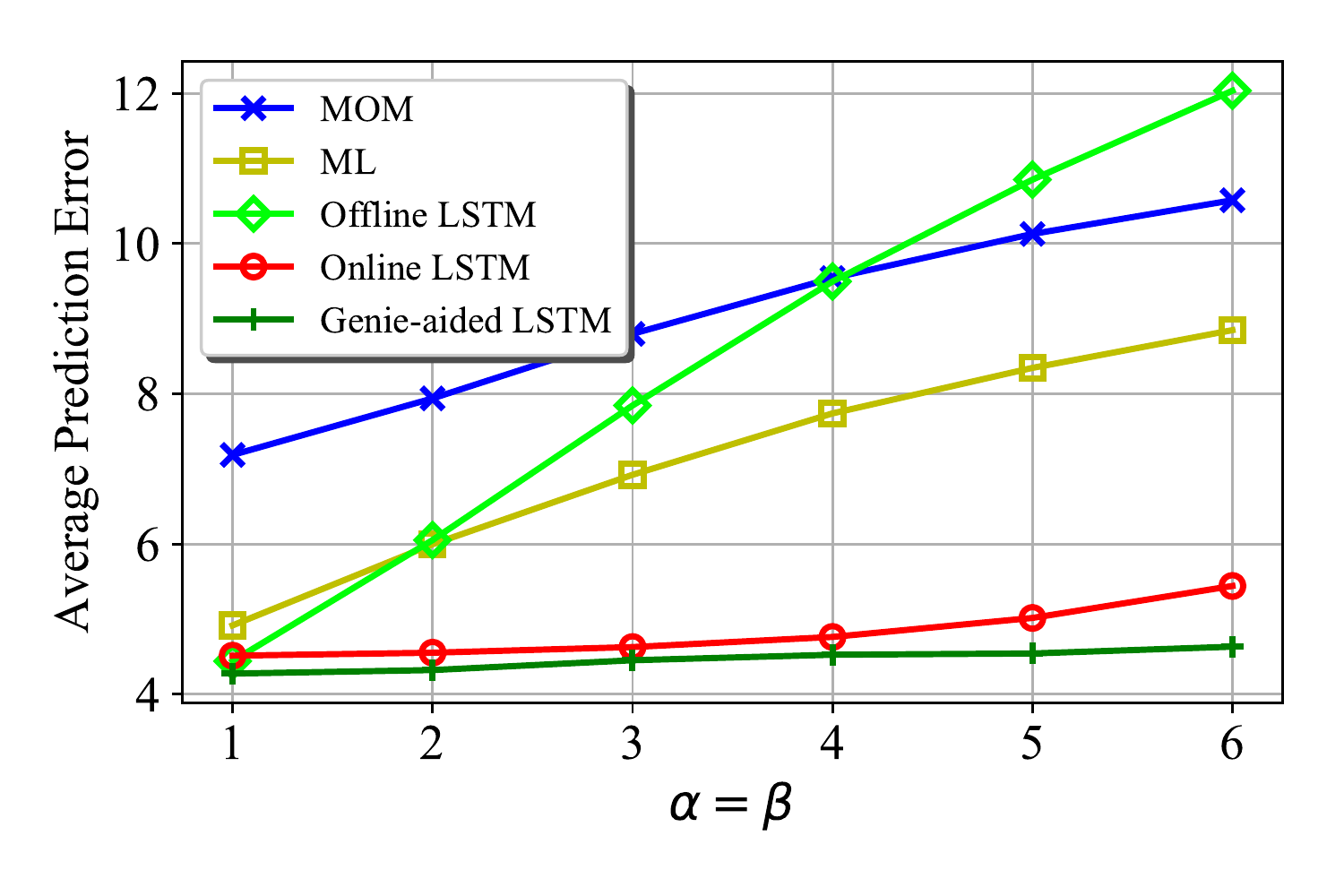}
        \vspace*{-0.8cm}
        \caption{\scriptsize Average prediction errors per frame versus the burstiness parameter ($\alpha,\beta$) for time limited Beta distributed traffic.}
                \label{fig7}
        \end{minipage}
    \end{center}
\vspace*{-0.5cm}
\end{figure}

We now consider a more general scenario, in which the $N_\text{p}= 100$ devices with periodic traffic generate packets at random according to the time-limited Beta profile \cite[Ch. 6.1]{3GPP2011Study} with period $T_\text{p}= 10$. The time-limited Beta profile defines a probability of packet generation that peaks in the middle of each period with burstiness defined by the magnitude of parameters ($\alpha,\beta$) \cite[Ch. 6.1]{3GPP2011Study}. Note that the deterministic packet generation considered so far is a special case of the current model, that is obtained by setting $\alpha=\beta\to\infty$. In Fig. \ref{fig7}, we plot the average prediction errors per frame versus a function of the burstiness parameter ($\alpha,\beta$). We observe that increasing ($\alpha,\beta$) degrades the prediction accuracy of all methods, since a higher burstiness results in a heavier traffic accumulation, which make prediction more difficult. The prediction accuracy of Offline LSTM is especially degraded, due to its lack of training observations under bursty traffic. In contrast, the proposed Online LSTM scheme only suffers from a minor performance degradation, demonstrating its capability to adapt to the traffic statistics.

\captionsetup{singlelinecheck=false} 

\vspace*{-0.1cm}
\section{Conclusions}
\vspace*{-0.0cm}
In this paper, we developed a traffic load prediction method based on online supervised learning in f-ALOHA networks. In the proposed method, LSTM RNNs are leveraged to capture temporal correlations due to traffic generation and protocol mechanisms. The scheme is based on a novel approximate labeling method based on backlog estimation. Numerical results demonstrate that the proposed method considerably outperforms conventional memoryless solutions, and that it can effectively adapt to new traffic statistics. A promising future direction is to develop lifelong learning and meta-learning techniques for online traffic prediction (see, e.g., \cite{park2019learning}).

\vspace*{-0.0cm}
\bibliographystyle{IEEEtran}
\bibliography{IEEEabrv,RA_bib}

\end{document}